\documentclass[english,aps,prb,twocolumn,superscriptaddress]{revtex4-1}
\usepackage[T1]{fontenc}
\usepackage[latin9]{inputenc}
\usepackage{amssymb}
\usepackage{graphicx}
\usepackage{esint}

\makeatletter
 
 \@ifundefined{textcolor}{}
 {%
   \definecolor{BLACK}{gray}{0}
   \definecolor{WHITE}{gray}{1}
   \definecolor{RED}{rgb}{1,0,0}
   \definecolor{GREEN}{rgb}{0,1,0}
   \definecolor{BLUE}{rgb}{0,0,1}
   \definecolor{CYAN}{cmyk}{1,0,0,0}
   \definecolor{MAGENTA}{cmyk}{0,1,0,0}
   \definecolor{YELLOW}{cmyk}{0,0,1,0}
 }

\makeatother

\usepackage{babel}
\begin{document}
\title{Signatures of localization in the effective metallic regime of high mobility Si MOSFETs}
\author{S. Das Sarma} 
\affiliation{Condensed Matter Theory Center and Joint Quantum Institute, University of Maryland, College Park, MD 20742, USA} 
\author{E. H. Hwang} 
\affiliation{Condensed Matter Theory Center and Joint Quantum Institute, University of Maryland, College Park, MD 20742, USA} 
\affiliation{SKKU Advanced Institute of Nanotechnology and Department of Physics, Sungkyunkwan University, Suwon, 440-746, Korea} 
\author{K. Kechedzhi} 
\affiliation{Condensed Matter Theory Center and Joint Quantum Institute, University of Maryland, College Park, MD 20742, USA} 
\author{L. A. Tracy} 
\affiliation{Sandia National Laboratories, Albuquerque, New Mexico 87185, USA} 
\begin{abstract}
Combining experimental data, numerical transport calculations, and theoretical analysis, we study the temperature-dependent resistivity of high-mobility two-dimensional 2D Si MOSFETs to search for signatures of weak localization induced quantum corrections in the effective metallic regime above the critical density of the so-called two-dimensional metal-insulator transition (2D MIT). The goal is to look for the effect of logarithmic insulating localization correction to the metallic temperature dependence in the 2D conductivity so as to distinguish between the 2D MIT being a true quantum phase transition versus being a finite-temperature crossover. We use the Boltzmann theory of resistivity including the temperature dependent screening effect on charged impurities in the system to fit the data. We analyze weak perpendicluar field magnetoresistance data taken in the vicinity of the transition and show that they are consistent with weak localization behavior in the strongly disordered regime $k_F\ell\gtrsim1$. Therefore we supplement the Botzmann transport theory with a logarithmic in temperature quantum weak localization correction and analyze the competition of the insulating temperature dependence of this correction with the metallic temperature dependence of the Boltzmann conductivity. Using this minimal theoretical model we find that the logarithmic insulating correction is masked by the metallic temperature dependence of the Botzmann resistivity and therefore the insulating $\ln T$ behavior may be apparent only at very low temperatures which are often beyond the range of temperatures accessible experimentally. Analyzing the low-$T$ experimental Si MOSFET transport data we identify signatures of the putative insulating behavior at low temperature and density in the effective metallic phase.
\end{abstract}
\maketitle

\section{Introduction\label{sec:Introduction}}

Si metal-oxide-semiconductor field effect transistors (Si MOSFETs)
support a highly conductive two dimensional electron gas (2DEG) system
where coupling to a metallic gate allows tuning of electron density
in the 2DEG over a wide range. Decreasing density can drive the 2DEG
from a highly conductive ``metallic'' state to a highly resistive
``insulating'' state~\cite{Ando_1982}. Intense theoretical and
experimental research over several decades suggested a number of possible
physical mechanisms underlying such behavior in 2DEGs depending on
microscopic details of the structure. The subject of the 2D metal-to-insulator
transition (2D MIT) arising from the gate-induced tuning of the 2D
carrier density is still an active area of research~\cite{Abrahams_2001,*Spivak_2010,*Sarma_Hwang_2005,*Kravchenko_Sarachik_2004},
particularly in the context of high-quality (i.e. high -mobility)
2D systems where the transition occurs at relatively low critical
density where electron-electron interaction effects may play an important
role. In particular, the specific question of whether the density-tuned
2D MIT is or is not a zero-temperature quantum phase transition, as
opposed to a crossover from a high-density apparent metallic phase
to a low-density insulating phase, has been much debated in the recent
literature~\cite{Abrahams_2001,*Spivak_2010,*Sarma_Hwang_2005,*Kravchenko_Sarachik_2004}.
If the 2D MIT turns out to be a true quantum phase transition rather
than a finite-temperature crossover, one immediate important implication
would be that the high-density 2D metallic phase must necessarily
be a non-Fermi liquid because a non-interacting 2D disordered Fermi
liquid is an insulator at $T=0$~\cite{Abrahams_2001,*Spivak_2010,*Sarma_Hwang_2005,*Kravchenko_Sarachik_2004, Abrahams_1979}.

Early measurements of 2D resistivity in low density Si MOSFETs showed
good agreement with the scaling theory of Anderson localization originating
from quantum interference of electrons scattered by random disorder
potential (see Refs.~\onlinecite{Ando_1982} and \onlinecite{Abrahams_2001,*Spivak_2010,*Sarma_Hwang_2005,*Kravchenko_Sarachik_2004}
and references therein). The theoretical argument~\cite{Abrahams_1979}
relies on the scaling theory of localization that shows that the system-size-independent
semi-classical Boltzmann (or Drude) resistivity $\rho_{B}$ in two
dimensions is overpowered by the logarithmic quantum correction $\sim\frac{1}{\pi}\ln\frac{L}{\ell}$
(in units of $h/e^{2}$) arising from quantum interference of diffusing
electrons, here $L$ is the system size and $\ell$ is the electron
mean free path. As a result, in the thermodynamic limit all states
are localized~\cite{Abrahams_1979} in a 2D orthogonal class system
(preserving time-reversal and spin-rotation symmetries). This result,
that all disordered 2D systems are insulating at $T=0$ in the infinite-system-size limit, was initially derived for non-interacting electrons,
but is universally thought to be valid in the presence of weak electron-electron
interaction. Boltzmann resistivity can be varied by tuning the electron
density in the 2DEG resulting in a tunable apparent metal-insulator
transition which occurs when the system size $L=\xi$ equals a characteristic
localization length at which the quantum correction equals the Boltzmann
part of the resistivity, $\rho_{B}\sim h/e^{2}$. This is of course
a finite-system-size induced crossover (and not really a transition)
from an effective apparent 2D metallic phase for $L\ll\xi$ to an
insulator for $L\gg\xi$.  Experimentally, however, the system-size
induced transition is impractical to implement, and one uses carrier
density to tune the effective localization length. This is possible
because the effective localization length $\xi\sim\ell\exp\left(\pi k_{F}\ell/2\right)$
depends on the underlying 2D carrier density through $k_{F}$ and
through the density-dependent mean free path $\ell$, and thus the
2D MIT can be tuned by changing the carrier density leading to a critical
density defining the crossover between the effective metal and the
strongly localized insulator.

In realistic experiments, there are inevitably inelastic scattering
processes that limit the coherent diffusion of electrons and introduce
a temperature-dependent dephasing length $L_{\varphi}\sim T^{-p/2}$.
This dephasing length limits the effective system size in the scaling
theory of localization providing a temperature-dependent length scale
cut off (i.e., the system size gets replaced by the dephasing length
in the scaling theory) which results in a logarithmic temperature
dependence of the quantum correction to resistivity. Therefore, in
a resistivity measurement, the metal-insulator transition is evidenced
by the qualitative change in the temperature dependence of resistivity
with changing electron density from a metallic (or, strictly speaking,
weak localization) dependence $\rho^{-1}(T)\approx\rho_{B}^{-1}-\mathrm{const}\times\ln T$
to an exponential insulating dependence $\rho(T)\sim\exp\left[\left(\frac{T_{0}}{T}\right)^{\alpha}\right]$,
characteristic of hopping or activated conduction, with some non-universal
scale $T_{0}$ and $\alpha=\frac{1}{3},\frac{1}{2},1$ depending on
the details. The presence of weak Coulomb interaction is not expected
to change the character of the quantum correction (at least from the
point of view of the perturbation theory in the high-density regime)
and only affects the coefficient in front of $\ln T$ due to the
additional scattering of electrons on Friedel fluctuations of density
around impurities~\cite{AAbook,AkkermansBook,Zala_2001,Zala_2001R,Gornyi_2004,Burdis_1988,Klimov_2008}
(the so-called Altshuler-Aronov effect).

By contrast, high-mobility Si MOSFETs $\mu\gtrsim20000\mathrm{cm^{2}/Vs}$
as well as a number of other high mobility 2DEGs seem to demonstrate
a qualitatively different dependence of resistivity $\rho(T,n)$ on
electron density $n$ and temperature~\cite{Abrahams_2001,*Spivak_2010,*Sarma_Hwang_2005,*Kravchenko_Sarachik_2004}.
In these samples, the low-density conductivity has the standard insulating
exponential temperature dependence. However, with increasing density
the temperature dependence of resistivity gradually changes from exponential
insulating behavior $d\rho/dT<0$ to a metallic-type $d\rho/dT>0$
dependence~\cite{Abrahams_2001,*Spivak_2010,*Sarma_Hwang_2005,*Kravchenko_Sarachik_2004,Zavaritskaya_1987,Vitkalov_2001,Tsui_2005,Kravchenko_1994,Kravchenko_1995,Kravchenko_1996,Pudalov_1998,Cham_1980,Smith_1986}
without any obvious manifestation of the $\ln T$ behavior on the
metallic side. Thus, there exists a range of densities where at the
lowest accessible temperatures only a metallic temperature dependence
is observed $d\rho/dT>0$. Actually, this metallic temperature dependence
(i.e., $d\rho/dT>0$) typically saturates at low temperatures ($T\lesssim100\mathrm{mK}$)
with the resistivity generically becoming temperature independent
(i.e., $d\rho/dT=0$) at low enough temperatures for all 2D effectively
metallic samples. Whether this low-temperature resistivity saturation
(with the actual value of the saturated low-temperature residual resistivity
being dependent on the carrier density) is a fundamental phenomenon
arising from some incipient low-energy cut-off suppressing the effective
metallicity or is just a trivial manifestation of electron heating
effect (where the carrier temperature saturates and no longer decreases
with the decreasing lattice temperature) is not known definitively.

In addition to this low-temperature resistivity saturation there is
a higher temperature anomaly in the metallic behavior also; typically,
the 2D metallic resistivity $\rho(T)$ starts decreasing (i.e. $d\rho/dT<0$)
at some density-dependent ``high'' temperature ($1-10\mathrm{K}$)
after manifesting the metallic (i.e. $d\rho/dT>0$) behavior and before
phonon scattering effects take over at still higher temperatures.
The combination of metallic (i.e. $d\rho/dT>0$) behavior at low temperatures
and insulating (i.e. $d\rho/dT<0$) behavior at intermediate temperatures
coupled with phonon-induced metallic behavior (i.e. $d\rho/dT>0$)
at still higher temperatures could lead to a rather interesting non-monotonicity
in $\rho(T)$ on the metallic side of the 2D MIT at low carrier densities,
and has been well-studied in the literature~\cite{DasSarma_Hwang_2000,*Min_Hwang_2012}.
The higher-temperature effective insulating behavior in the metallic
phase is thought to arise from a quantum-classical crossover phenomenon
in the 2D system occurring on the scale of the Fermi temperature ($\propto n$)
which could be low ($\lesssim10\mathrm{K}$) at the low carrier densities
of interest for the 2D MIT phenomena~\cite{DasSarma_2003}. We will
not much discuss this temperature-induced quantum-classical high-temperature
transition from metallic to insulating behavior in this paper, concentrating
instead on the density-induced 2D MIT transition at low temperatures.
The sign of the derivative switches from insulating to metallic at
a value of resistivity of the order of the resistance quantum $\rho\sim h/e^{2}$,
i.e. at the value at which a transition to strong localization behavior
is predicted by the scaling theory. On the metallic side of this transition,
where $d\rho/dT>0$, the resistivity increases sharply by a factor
of $\gtrsim2-3$ with growing temperature at lower carrier density
staying within the metallic phase. This pronounced temperature dependence
diminishes with growing density deeper in the metallic regime. Standard
$\ln T$ quantum corrections are observed at high densities where
the metallic temperature dependence is weak~\cite{Pudalov_1998logT,Klimov_2008,Pudalov_1999}.
By contrast in the vicinity of the metal-insulator transition logarithmic
corrections are typically not observed within the experimentally accessible
temperature range. This qualitative change in the temperature dependence
of 2D resistivity driven by electron density is routinely called a
metal-insulator transition in the literature and we will use this
convention in the following despite the ongoing debate about the existence
or not of an actual thermodynamic phase transition at this point~\cite{Abrahams_2001,*Spivak_2010,*Sarma_Hwang_2005,*Kravchenko_Sarachik_2004}.
Our view in the current work is based on the assumption that the 2D
MIT is a crossover phenomenon with the $\ln T$ behavior suppressed
by the strong metallic temperature dependence of the Drude-Boltzmann
resistivity at lower metallic densities. We will critically test this
assumption in this paper by comparing theory and experiment in the
density- and temperature-dependent transport data in high-mobility
Si MOSFETs.

An important distinction between low- and high-mobility samples is
in the relative strength of Coulomb interactions. The presence of
weaker disorder high mobility structures allows for metallic behavior
to persist down to very low densities $n\sim10^{11}cm^{-2}$ which
correspond to very small values of Fermi energy and thus large values
of the density dependent dimensionless interaction strength in the
system which may be as large as $r_{s}\equiv1/\sqrt{\pi na_{B}^{2}}\sim10$.
Here $a_{B}=\hbar\kappa/(me^{2})$ is the effective Bohr radius of
electrons in the 2DEG, $\kappa$ being the background dielectric constant.
We note, however, that even high-density Si MOSFETs, which were extensively
studied~\cite{Ando_1982} before the current interest arose in the
2D MIT phenomena, have a dimensionless interaction strength $r_{s}>1$,
and 3D metals all have $r_{s}\sim4-7$. Thus, it is not manifestly
clear that interaction by itself is the sole physical mechanism underlying
the 2D MIT phenomena. Perhaps an even more important aspect of high-mobility 2D samples in the context of strong metallic temperature
dependence of resistivity is that, by having a relatively low critical
density distinguishing the metallic and the strongly insulating regimes
by virtue of low sample disorder (and hence high sample mobility),
the Fermi temperature ($T_{F}\propto n$) is low ($\sim1-10\mathrm{K}$)
in high mobility samples in the metallic regime. Then, the low-temperature
regime ($T\approx0.1-5\mathrm{K}$) where the 2D metallic behavior
manifests itself (i.e. large positive $d\rho/dT$) has effectively
large values of the dimensionless temperature $T/T_{F}\sim1$. By
contrast, 3D metals, which are also strongly interacting electron
systems by virtue of having $r_{s}\gg1$ have very low dimensionless
effective temperature $T/T_{F}\sim10^{-4}$, by virtue of $T_{F}\sim10^{4}\mathrm{K}$
in metals. The large effective values of $T/T_{F}$ also distinguish
the high-mobility 2D systems from the low-mobility 2D systems where
the Fermi temperature is $T_{F}\gtrsim100K$ in the metallic phase
and thus $T/T_{F}\sim10^{-2}-10^{-3}$ in the low-temperature experimental
regime.

Extensive theoretical work demonstrated that all of the observed features
of the temperature dependence of resistivity on the metallic side
of the metal-insulator transition can be successfully described extrapolating
from high densities (and coincidentally low interactions strength)
and using a Boltzmann transport theory which includes the temperature
dependent screening~\cite{DasSarma_1999,DasSarma_2003,DasSarma_2004}
of charged impurities. The metallic increase of the resistivity with
growing temperature is explained by a decreasing efficiency of screening
by the electron gas of charged impurities with growing temperature.
The large effective value of $T/T_{F}$ explains the experimentally
observed large $d\rho/dT$ in the metallic phase. This suggests that
the standard Fermi liquid theory may explain the unusual temperature
and density dependence of resistivity in high mobility Si MOSFETs.
However, the Fermi liquid theory also predicts the presence of quantum
corrections giving rise to $\ln T$ behavior which are not observed
experimentally in the vicinity of the metal-to-insulator transition.
The observation and analysis of these $\ln T$ corrections would
allow us to continuously connect the low-density strong interaction and
strong disorder regime to the weakly interacting high-density Fermi
liquid regime where the Boltzmann theory is valid. There may be a
simple conventional explanation for the absence of quantum corrections
in the data. Analyses of the high-density data~\cite{Pudalov_1999,Pudalov_1998logT,Altshuler_Martin_Maslov_Pudalov_Prinz_Brunthaler_Bauer_2000,Altshuler_2001}
where $\ln T$ is observed suggest that at low densities in high-mobility
samples the temperature at which $\ln T$ would become apparent may
be beyond the measurement temperatures because of electron heating~\cite{Altshuler_2001,Prus_2001}.
Our approach in this paper is a straightforward phenomenological
approach where we assume that the metallic transport has contributions
from both the screening-induced semi-classical metallic temperature
dependence and the weak localization induced quantum $\ln T$ temperature
dependence. The strong metallic temperature dependence of the resistivity
completely overwhelms the $\ln T$ insulating correction at higher
temperatures with the logarithmic correction eventually manifesting
itself at some density-dependent low temperatures which might very
well be inaccessible to experimental measurements due to electron
heating problem. We also present experimental transport data on 2D
MIT in Si MOSFET samples which are consistent with the presence of
both screening-induced metallic temperature dependence and quantum
weak localization correction.

In this paper, we present experimental resistivity data taken on two
high-mobility Si MOSFETs demonstrating 2D metal-insulator transition.
We construct a microscopic Boltzmann theory of resistivity in Si MOSFETs
that includes the effect of temperature-dependent screening of charged
impurities by the electron gas. We use this model to fit the metallic
temperature dependence observed in the data. We then construct a minimal
additive model describing the competition between this metallic temperature
dependence of resistivity and the insulating temperature dependence
due to the quantum correction. Using this model, we determine the temperatures
at which the quantum correction to resistivity is expected to dominate
the experimental data which turn out to be beyond the range of the
current experiments. We also discuss the magnetoresistance data on
the two samples in the vicinity of the transition and show that they
are qualitatively consistent with weakly interacting localization
theory suggesting that the standard Fermi-liquid theory could be sufficient
to describe the temperature dependence of the transport properties
in high-mobility Si MOSFETs.

\section{Description of the experiment}

\begin{figure}
\includegraphics[width=0.5\textwidth]{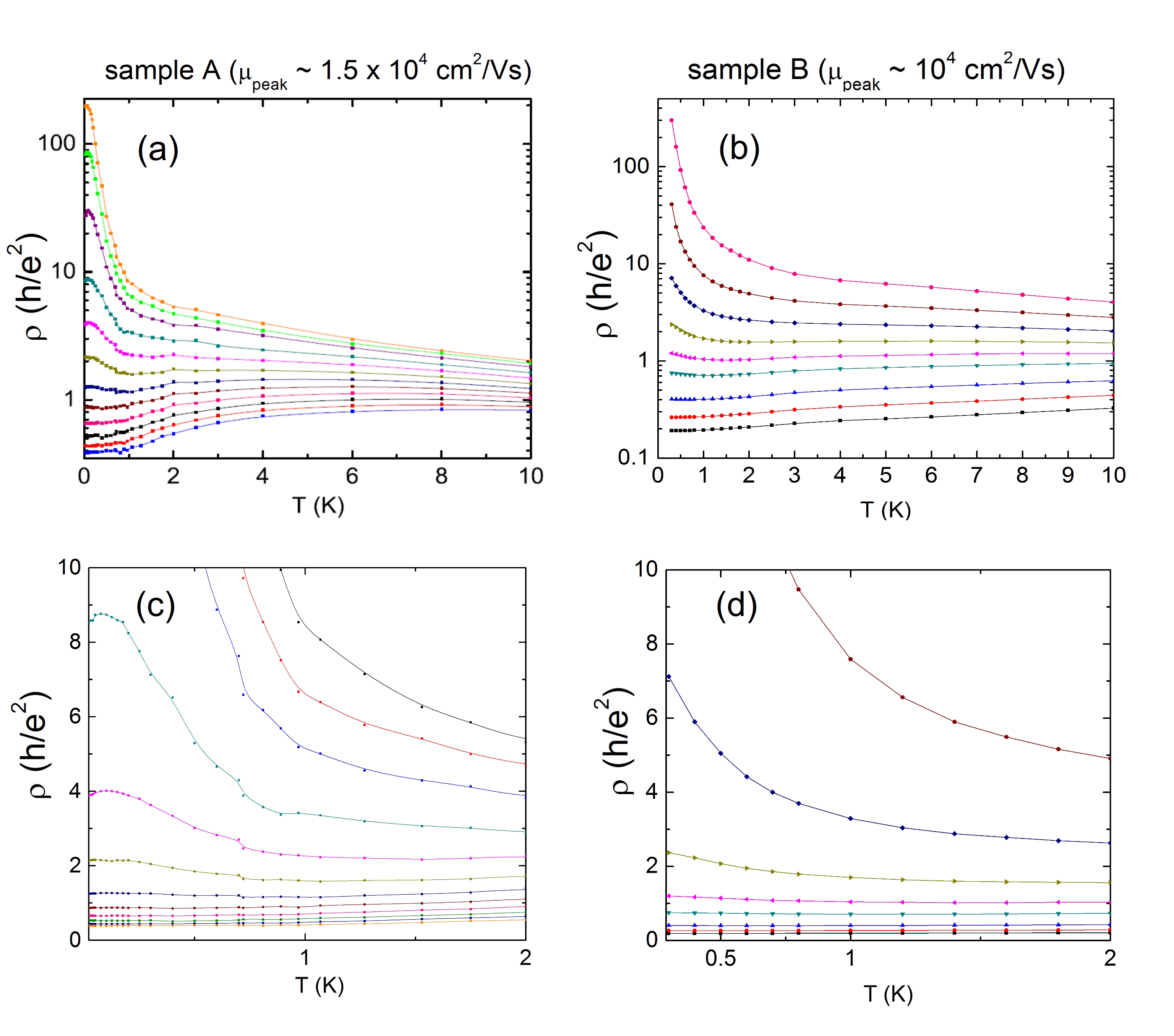}\caption{Resistivity as a function of temperature for sample A and B plots
((a) and (c)), and ((b) and (d)), respectively showing zoom in of
the low-temperature region in (c) and (d). Different lines correspond to different carrier density $n$ in units of $10^{11}\mathrm{cm^{-2}}$
(from top to bottom): in (a) and (c) $n=1.07,\,1.10,\,1.13,\,1.20,\,1.26,\,1.32,\,1.38,\,1.44,\,1.50,\,1.56,\,1.62,$
and $1.68$; in (b) and (d) $n=1.52,\,1.70,\,1.88,\,2.05,\,2.23,\,2.41,\,2.76,\,3.11,$
and $3.46$. Plot (a) is reproduced from Ref.~\onlinecite{Tracy_2009}.}
\label{rhoT}
\end{figure}

We consider the transport data on two Si MOSFET samples with relatively
high mobility: sample A ($\mu=1.5\times10^{4}\mathrm{cm^{2}/Vs}$)
and sample B ($\mu=10^{4}\mathrm{cm^{2}/Vs}$). The resistivity temperature
dependence at various carrier densities is shown in Fig.~\ref{rhoT}.
At the lowest densities an insulating exponential temperature dependence
is observed. This insulating behavior $d\rho/dT<0$ gradually loses
exponential character with increasing density. At higher densities
the temperature dependence of resistivity becomes non-monotonic: with
increasing temperature the resistivity drops down to a minimum value
and then rises up $d\rho/dT>0$ to a maximum before gradually sloping
downwards. There is a range of densities at which at the lowest accessible
temperatures there is no sign of the insulating rise of the resistance
or $\ln T$ metallic correction. This form of temperature dependence
$\rho(T)$ is typical in high-mobility Si MOSFETs~\cite{Zavaritskaya_1987,Vitkalov_2001,Tsui_2005,Kravchenko_1994,Kravchenko_1995,Kravchenko_1996,Pudalov_1998,Cham_1980}.
We point out, as is obvious from Figs.~\ref{rhoT}~(c) and~(d) where
the resistivity is shown on an expanded temperature scale, the insulating
temperature dependence is suppressed gradually as density increases
and there is really no absolutely sharp density distinguishing metallic
and insulating behaviors.

\begin{figure}
\includegraphics[width=0.5\textwidth]{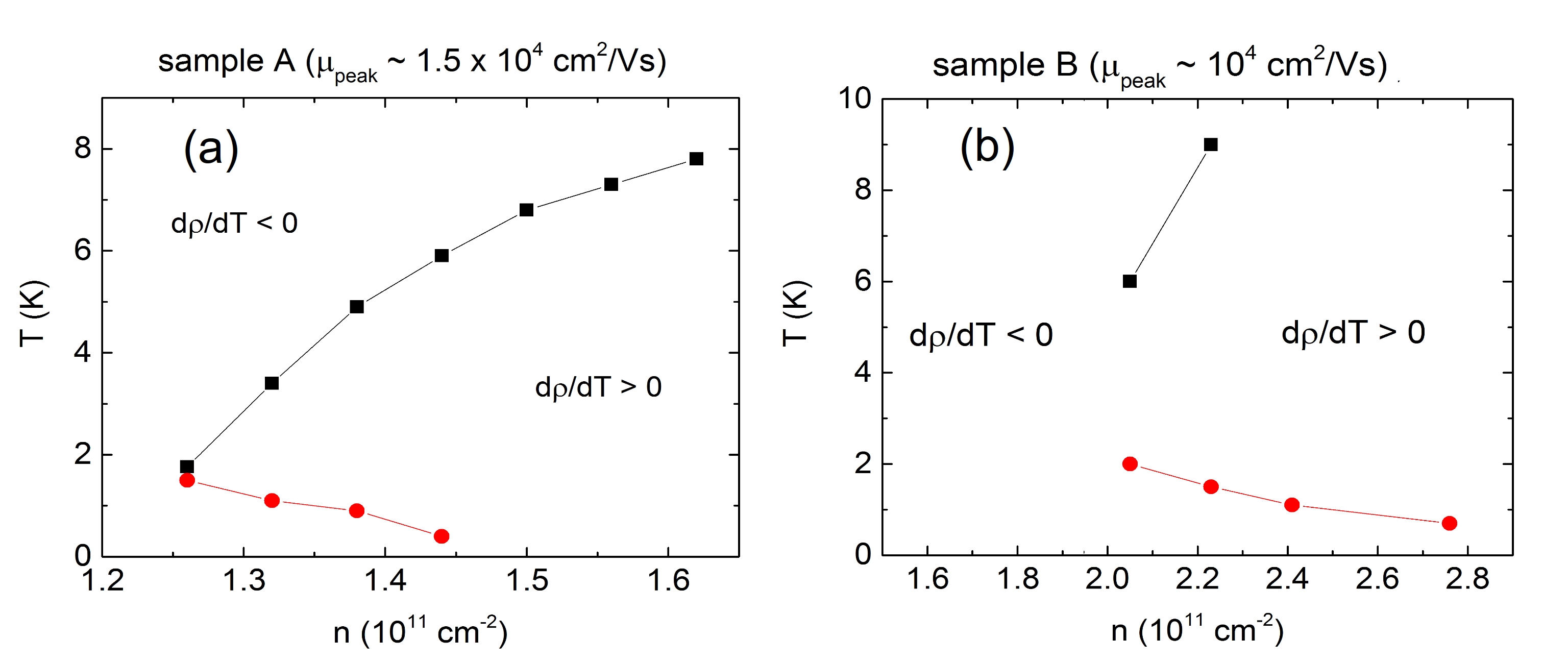}\caption{Black squares and red circles correspond to the maximum and minimum
of the measured resistivity $\rho(T)$ in Fig.~\ref{rhoT}, respectively. }

\label{MinMax}
\end{figure}

In Fig.~\ref{MinMax}, we track the evolution with electron density
of the temperatures at which the maximum and minimum of the resistivity
are reached. Red circles here correspond to the low-temperature resistivity
minimum which signifies the onset of insulating behavior. The black
squares correspond to the high-temperature maximum of the resistivity.
Our interest in this work is mostly on the red circles which
provide the temperature at which the metallic temperature dependence
is just being overcome by the quantum localization effect on the effective
metallic side of the 2D MIT. We note that as expected this characteristic
temperature for the resistivity minimum is rather low and it increases
with decreasing density as localization effects become more important
quantitatively. The black squares in Fig.~\ref{MinMax}, indicating
the resistivity maxima in $\rho(T)$ as a function of carrier density,
provide the characteristic temperature for the high-temperature quantum-to-classical
 crossover in the 2D transport as discussed in Sec.~\ref{sec:Introduction}.
This quantum-classical crossover typically occurs on the scale of
the Fermi temperature (typically around $T\sim T_{F}/2$) and therefore
decreases with decreasing carrier density. The region in-between the
red circles and the black squares is the putative 2D effective metallic
phase where the metallic temperature dependence with $d\rho/dT>0$
is manifested in 2D transport. We note that in the high-mobility samples,
the regime (below red circles) showing $\ln T$ weak localization
behavior is strongly suppressed by the metallic temperature dependence
arising from other physical mechanisms.

\section{Boltzmann theory}\label{sec:Boltzmann}

Strong disorder and strong interaction low-density regime is difficult
to address theoretically. Boltzmann theory allows us to quantitatively
describe mobility dependence on the electron density in a wide density
regime (of not too low densities so that one is away from the strongly
localized regime). The model of disorder describing the density dependence
of mobility is a combination of random charged impurities and surface
roughness~\cite{Ando_1982}. This Boltzmann theory quantitatively
agrees with mobility measurements in a wide density range in the higher-density
 metallic regime~\cite{Tracy_2009}. It is therefore natural
to extrapolate this theory to the low-density regime of strong disorder
and interaction. At very low densities, the effect of surface roughness
is negligible and the resistivity is completely dominated by charged
impurities~\cite{Ando_1982}. In the following, we neglect the effect
of surface roughness, but our results do not change if surface roughness
is included in the theory since it has little quantitative effect
on transport at the low carrier densities of interest in this
work. 

Boltzmann conductivity $\sigma_{B}$ is given by,

\begin{equation}
\rho_{B}^{-1}\equiv\sigma_{B}=ne^{2}\left\langle \tau\right\rangle /m\label{eq:Boltzmann}
\end{equation}
where $n$ is the electron density, $m$ is the effective mass of
electrons, and the average transport time $\langle\tau\rangle$ reads as,

\begin{equation}
\left\langle \tau\right\rangle \equiv\frac{\int dE\tau(E)E\left(-\frac{\partial f}{\partial E}\right)}{\int dEE\left(-\frac{\partial f}{\partial E}\right)}.
\end{equation}
Here the impurity averaged relaxation time $\tau(E)$ is given by

\begin{eqnarray}
 & \frac{1}{\tau(E_{k})}=\frac{2\pi}{\hbar}\sum_{\alpha,\mathbf{k'}}\int_{-\infty}^{\infty}dzN_{i}(z)|u(\mathbf{k}-\mathbf{k'};z)|^{2}\nonumber \\
 & \times(1-\cos\theta_{\mathbf{kk'}})\delta(E_{k}-E_{k'}),
\end{eqnarray}
where the standard parabolic energy dispersion is assumed for $E_{k}$,
$N_{i}(z)$ is the 3D charged impurity density and $u(\mathbf{q};z)$
is the 2D Fourier transform in the plane of the 2DEG of the impurity
potential screened by the electron gas,

\begin{equation}
u(q;z)=\frac{1}{\varepsilon(q)}\frac{2\pi e^{2}}{\kappa q}F_{imp}(q;z),
\end{equation}
where the $\kappa$ is the background dielectric constant, and $F_{imp}(q;z)$
is a form factor depending on the microscopic details which are known~\cite{Ando_1982}.
In the strictly 2D limit an impurity charge located a distance $d$
away from the 2DEG is described by a form factor $F_{imp}=e^{-qd}$.
The screening effect is characterized by the dielectric function $\varepsilon(q)$
calculated in the random phase approximation (RPA)~\cite{Ando_1982}.

\begin{figure}
\includegraphics[clip,width=0.5\textwidth]{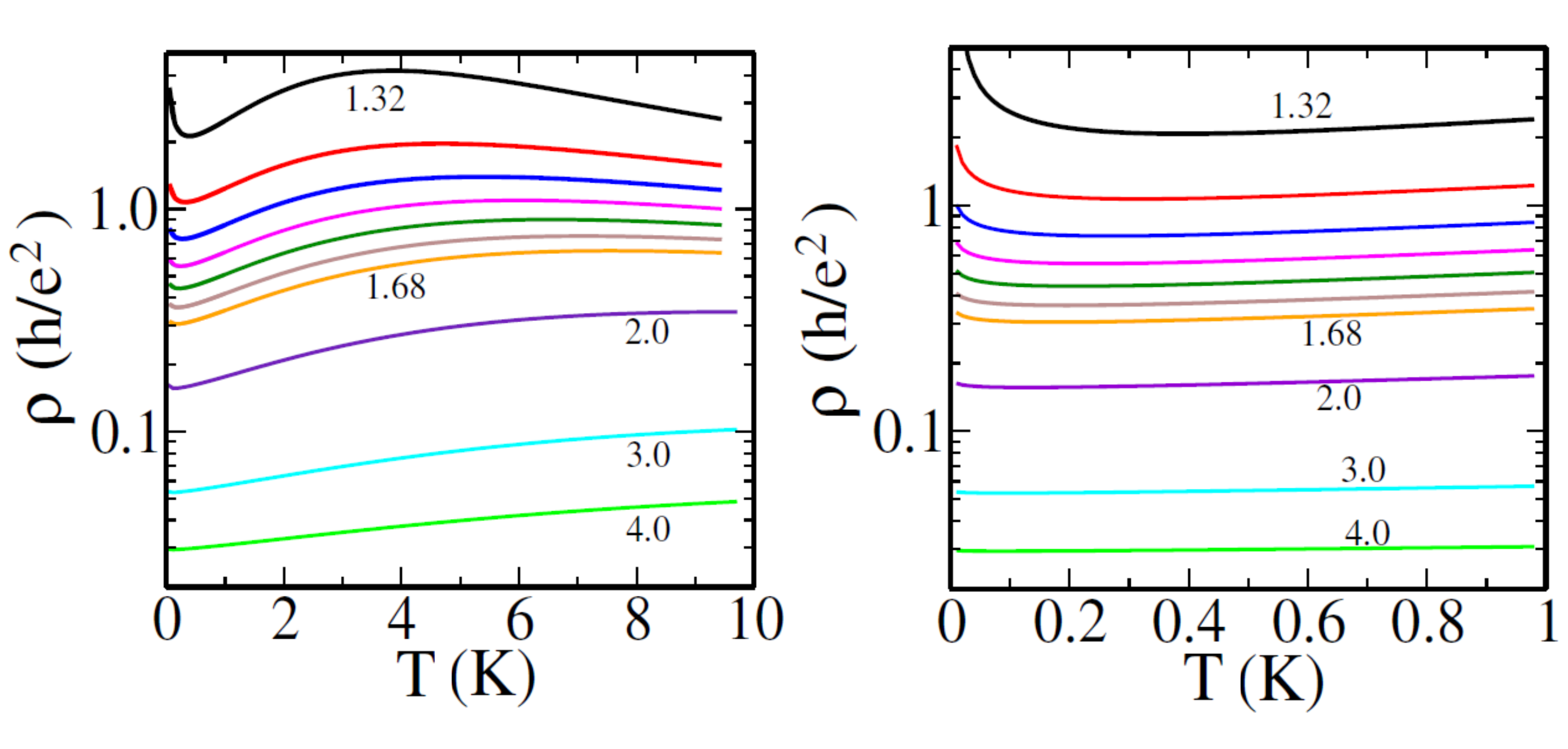}

\caption{Boltzmann resistivity theory for parameters of Sample A (from fitting)
combined with the weak localization correction Eq.~(\ref{eq:Resistivity}).
Numbers on the plot correspond to the electron density in units of
$10^{11}cm^{-2}$. The right panel shows a zoom in on a low-temperature
region in the left panel. }

\label{rhoTtheory}
\end{figure}

The resulting theory fits well (see Fig.~\ref{rhoTtheory}) the strong
non-monotonic temperature dependence of the metallic resistivity at
low densities~\cite{DasSarma_1999,DasSarma_2003}. To reproduce the experimental data,
we adjust the density of impurities and their locations in the oxide layer in a narrow range of values as free fitting parameters and the resulting model reproduces the functional dependence of the resistivity on temperature and electron density in the 2DEG. We mention that the oxide impurity charge density and their spatial distribution necessary to get agreement between the theory and the experimental resistivity data are very reasonable (and independently confirmed by capacitance measurements).  The details of this comparison between theory and experiment for our samples are given in Ref.~\onlinecite{Tracy_2009} and not repeated here. The non-monotonicity of the temperature dependence of resistivity as a function of density can therefore be explained within this model: with increasing temperature screening becomes less effective and as a result the resistivity
increases. The increase in resistivity can be by a factor $2-3$ in
this regime due to the temperature-induced weakening of the screening
effect which explains the observed metallic temperature dependence.
At higher temperatures, the quantum-to-classical crossover results
in the decreasing resistivity with temperature at $T/T_{F}\sim1$. The theoretical details of the screening model for the 2D resistivity and the corresponding comparisons with experimental "metallic" temperature dependence of transport properties have already been discussed extensively in earlier references~\onlinecite{DasSarma_Hwang_2000,*Min_Hwang_2012,DasSarma_2003,DasSarma_1999,DasSarma_2004,Tracy_2009} and will not be repeated here.  We mention, however, that none of these earlier references included the weak localization effect into consideration (as we do in this work) assuming the effective metallic behavior to dominate the transport properties completely.

Extrapolation to the strong interaction regime makes sense since the
random phase approximation (RPA) is given by a subset of the most
divergent diagrams which therefore may dominate even at strong interactions~\cite{DasSarma_1999,DasSarma_2003}.
Extrapolation of Boltzmann theory to low density and strong interaction
may be successful as it describes short-range phenomena $\lesssim\ell$
(where $\ell$ is the mean free path) as opposed to localization physics
and localizing interaction correction originating from the diffusive
length scales $\mbox{\ensuremath{\gg\ell}}$. Boltzmann theory therefore
may describe the case of strong dephasing and/or high temperature.
In particular, the fact that the effective temperature is high (i.e.
$T/T_{F}\sim1$) in the low-density high-mobility samples makes 2D
MIT a high-temperature crossover phenomenon where interaction effects
are likely to be strongly suppressed by temperature. We note here
(as can be seen in Fig.~\ref{rhoT}) that at very low temperatures,
the temperature dependence of the metallic resistivity invariably
saturates, thus making the 2D MIT an effectively high-temperature
phenomenon.

\section{Magnetoresistance}

A strong indication of a Fermi liquid behavior in the strong disorder
and strong interaction regime near the metal-insulator transition
is the observation of weak perpendicular field magnetoresistance which
is a smoking gun signature of the weak localization physics. Magnetoresistance
data taken on sample A (see Fig.~5 in Ref.~\onlinecite{Tracy_2009})
were fitted using the standard di-gamma function expression for the
weak localization theory,

\begin{equation}
\frac{\delta\rho}{\rho^{2}}=-\alpha g_{v}G_{0}\left[\Psi\left(\frac{1}{2}+\frac{\tau_{B}}{\tau}\right)-\Psi\left(\frac{1}{2}+\frac{\tau_{B}}{\tau_{\varphi}}\right)\right],\label{eq:WLMR}
\end{equation}
where $\tau_{B}\equiv\hbar/(4eBD)$, $G_{0}=\frac{e^{2}}{2\pi^{2}\hbar}$,
$B$ stands for the perpendicular magnetic field, $D$ the diffusion
coefficient, $g_{v}$ the valley degeneracy factor. The coefficient
$\alpha$ along with the dephasing time $\tau_{\varphi}$ are used
as fitting parameters with the best fit achieved with $\alpha\approx0.25$
and $\tau_{\varphi}=33,32,$ and $28ps$ for electron densities $n=1.45,1.51,$
and $1.63\times10^{11}cm^{-2}$ at $T=0.1K$ ($g_{v}=1$ is assumed
in the fits).

In this low density regime of these measurements the resistivity is
high $\rho\lesssim h/e^{2}$, which suggests $k_{F}\ell\gtrsim1$,
whereas the standard weak-localization theory is really valid for
$k_{F}\ell\gtrsim10$. Therefore extra care has to be taken when interpreting
these results and quantum corrections of higher order in $1/(k_{F}\ell)$
may have to be considered going beyond the usual weak localization
theory. The key effect of the higher order terms is in lowering the
prefactor in front of the magnetoresistance expression Eq.~(\ref{eq:WLMR})
from $\alpha=1$ to $\alpha\approx1-\beta(G_{0}\rho)$, with $\beta$
a degeneracy factor depending on the intervalley scattering in the
system. This reduction in $\alpha$ is a result of the two-loop correction
to non-interacting weak localization theory~\cite{Minkov_2004}.
Also an additional magnetoresistance due to electron-electron interactions
may enhance the reduction of the prefactor with the combined effect
leading to $\alpha\approx1-2\beta(G_{0}\rho)$, as discussed in Ref.~\onlinecite{Minkov_2004}. 

Intervalley scattering due to the short range scattering may suppress
the valley degeneracy factor $g_{v}$ in front of the magnetoresistance
in Eq.~(\ref{eq:WLMR}). The effect of intervalley scattering on
magnetoresistance was analyzed in great detail using high density
measurements in high mobility Si MOSFETs in Ref.~\onlinecite{Gershenson_2007}.
The typical intervalley scattering times extracted from these measurements
are in the range $1ps\lesssim\tau_{v}\lesssim20ps$. Comparing the
typical values of $\tau_{v}$ with the dephasing time $\tau_{\varphi}\sim30ps$
in our samples extracted from our measurements we conclude that $\tau_{v}/\tau_{\varphi}\lesssim1$
and valley mixing is relatively strong and the effective degeneracy
factor is expected to be in the range $1\lesssim g_{v}$ in Eq.~(\ref{eq:WLMR}).
This means that the fitting parameter $\alpha g_{v}=0.5$ in Eq.~(\ref{eq:WLMR})
signifies a reduction of the magnetoresistance amplitude by at least
a factor of two, and probably more. This suggests that the data in
our Si MOSFETs is qualitatively consistent with the detailed theory~\cite{Minkov_2004}
of quantum corrections in a weakly interacting strongly disordered
Fermi liquid in the presence of strong inter-valley scattering (which
may arise from the surface roughness at the $\mathrm{Si-SiO_{2}}$
interface providing short-range scattering). This agreement has to
be taken with a grain of salt as the interaction strength in the low
density regime may not be small. Nevertheless, similar magnetoresistance
behavior was observed in other high mobility Si MOSFET measurements~\cite{Brunthaler_2001}
and other 2DEGs~\cite{Simmons_1998,Coleridge_2002,Rahimi_2003} giving
us confidence in this conclusion. 

Experimental values of the magneto-resistance cut off extracted from
the fitting procedure give a dephasing time $\tau_{\varphi}(0.1K)\sim30ps$
in our sample which is an order of magnitude shorter than the value
expected due to inelastic electron-electron scattering in the diffusive
regime~\cite{AAbook,AkkermansBook,Narozhny_2002},

\begin{equation}
\frac{\tau}{\tau_{\varphi}}=\gamma\frac{k_{B}T}{E_{F}}\ln g,\label{eq:tauPhi}
\end{equation}
where $\gamma$ is a factor of the order unity. Despite the low-mobility
2D devices showing quantitative agreement with the dephasing rate
due to inelastic electron-electron interaction~\cite{Davies_1983,Kawaji_1986}
in Eq.~(\ref{eq:tauPhi}), and also high-density high-mobility devices
showing quantitative agreement with the dephasing rate formula~\cite{Gershenson_2007},
low density measurements routinely manifest an order of magnitude
shorter dephasing times than predicted by Eq.~(\ref{eq:tauPhi})~\cite{Brunthaler_2001},
which cannot be explained by simple deviations~\cite{Minkov_2004}
from Eq.~(\ref{eq:tauPhi}) at strong disorder $k_{F}\ell\sim1$.
The same enhanced dephasing rate is also routinely observed in other
2DEGs~\cite{Simmons_1998,Coleridge_2002,Rahimi_2003}. It seems likely
that either there is an additional dephasing mechanism responsible
for such short dephasing rates or the magnetoresistance is cut off
by the localization length. At low values of $k_{F}\ell\gtrsim1$
the localization length $\xi\approx\ell\exp\left(\pi k_{F}\ell/2\right)$
becomes comparable to the dephasing length. It has been shown theoretically
that Eq.~(\ref{eq:WLMR}) is applicable even for $\xi<L_{\varphi}$
as long as $k_{F}\ell>1$. However, the meaning of the dephasing rate
extracted from the data is different in this regime since the localization
length cuts off the magneto-resistivity instead of dephasing~\cite{Minkov_2004}.
This may cause a saturation in the temperature dependence of the dephasing
rate. It is, in principle, also possible that the dephasing rate at
very low carrier density is dominated by the physics of density inhomogeneity~\cite{Germanenko_2001}
(i.e. disorder induced puddles) not included in the theory leading
to Eq.~(\ref{eq:tauPhi}).

\section{Temperature dependence of resistivity within the Fermi liquid model\label{sec:Temperature-dependence-of}}

\begin{figure}
\includegraphics[width=0.3\textwidth]{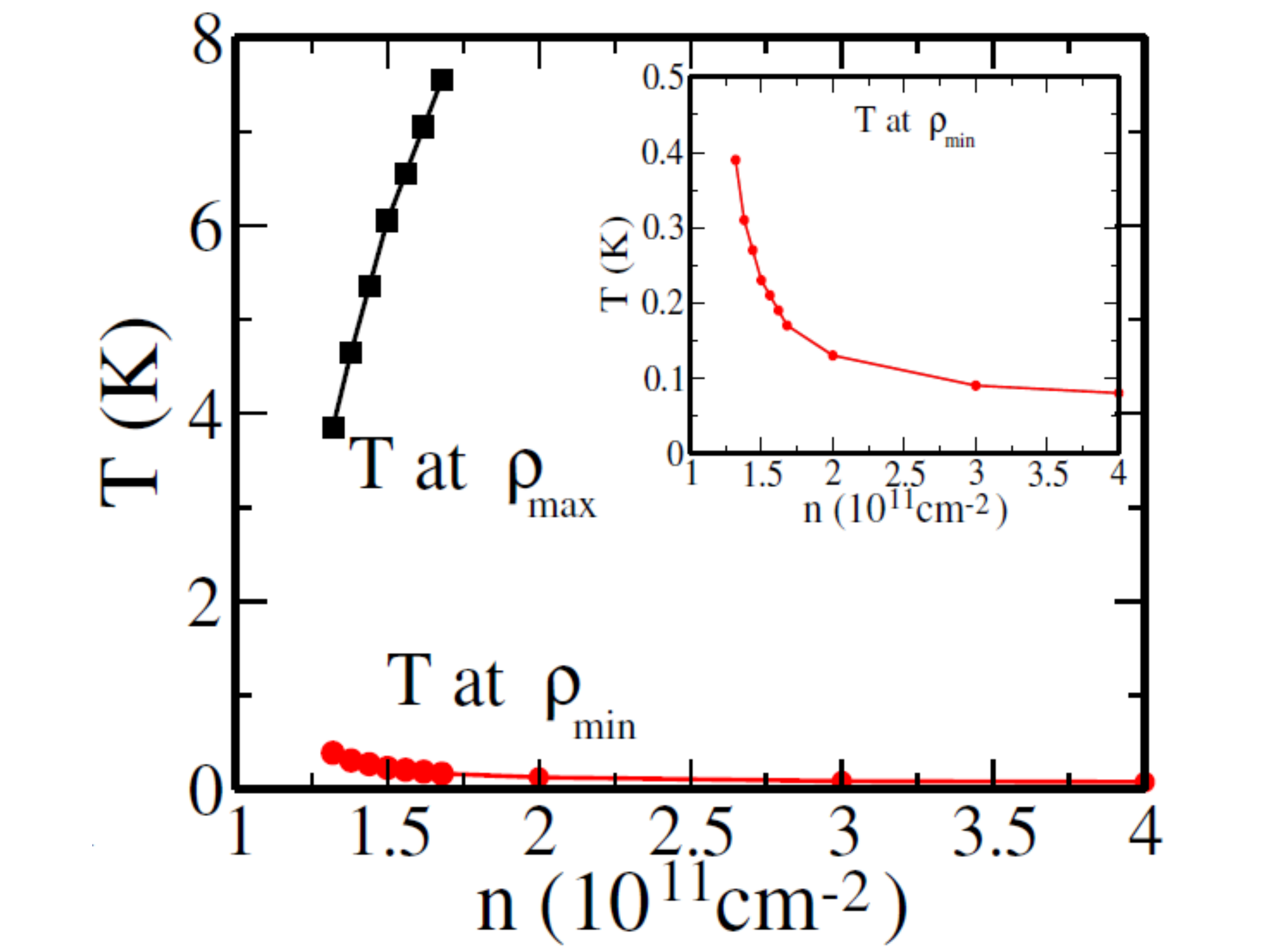}\caption{Black squares and red circles correspond to maximum and minimum of
$\rho(T)$ in Eq.~(\ref{eq:Resistivity}). Inset shows temperature
at the minimal value of resistance as a function of electron density.}

\label{fig:MinMaxTheory}
\end{figure}

High-density measurements identified standard $\ln T$ quantum correction
to 2D resistivity~\cite{Pudalov_1998logT,Brunthaler_2001,Pudalov_1999}.
It is therefore natural to expect that, since the Boltzmann theory
can be continuously extended to low densities, the $\ln T$ correction
is present in the system at all densities in addition to the Boltzmann
contribution. The presence of weak field magnetoresistance discussed
above is an additional argument in favor of the Fermi liquid behavior
at low densities near the metal-insulator transition. Therefore, we
assume the presence of $\ln T$ quantum correction to the Boltzmann
resistivity up to the onset of the strongly insulating behavior~\cite{Pudalov_1998logT,Altshuler_Martin_Maslov_Pudalov_Prinz_Brunthaler_Bauer_2000,Altshuler_2001}.
However, with decreasing electron density, the metallic temperature
dependence of the Boltzmann resistivity becomes pronounced. As a
result, there is a competition between the insulating and metallic
temperature dependence at low densities which are simultaneously present
since they arise from distinct microscopic mechanisms. We consider
the minimal model of transport that describes this behavior, including
both semi-classical Boltzmann contribution and the quantum weak localization
contribution,

\begin{equation}
\rho(T)=\frac{1}{\rho_{B}^{-1}(T)+\sigma_{WL}(T)},\label{eq:Resistivity}
\end{equation}
where $\rho_{B}(T)$ is the temperature-dependent Boltzmann resistivity
given by Eq.~(\ref{eq:Boltzmann}). The quantum correction to conductivity
in Eq.~(\ref{eq:Resistivity}) is given by the standard theory,

\begin{equation}
\sigma_{WL}=-g_{v}G_{0}\frac{1}{2\pi}\ln\frac{\tau_{\varphi}}{\tau}.\label{eq:WLT0}
\end{equation}

Figure~\ref{rhoTtheory} shows the calculated temperature and density
dependence of the resistivity for the parameters extracted from the
fits of the data for sample A. In Fig.~\ref{fig:MinMaxTheory}, we
present the theoretical results corresponding to those shown in Fig.~\ref{MinMax}.
It is clear that the theoretical results in Fig.~\ref{fig:MinMaxTheory}
closely resemble the experimental results shown in Fig.~\ref{MinMax}
for sample A, thus demonstrating that 2D MIT may indeed be a crossover
phenomenon. Here the logarithmic correction becomes apparent only
below the experimentally accessible temperatures $T\sim0.1\textrm{K}$ which
is qualitatively similar to the experimental situation.

\begin{figure}
\includegraphics[width=0.5\textwidth]{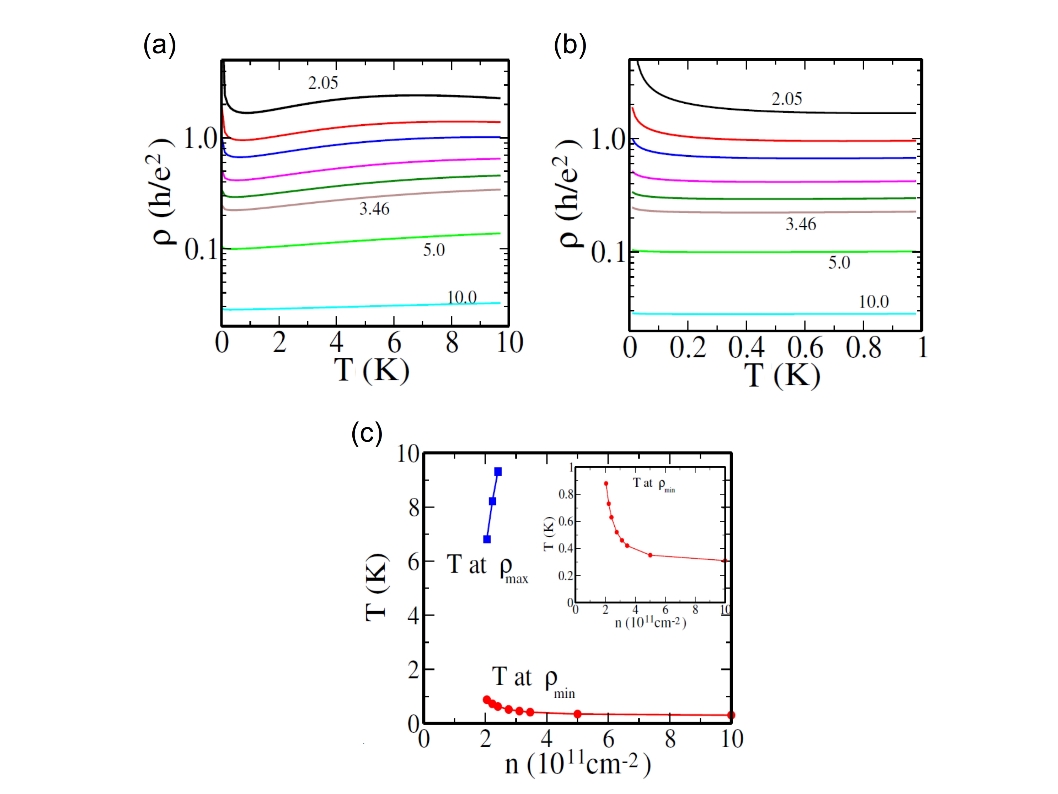}\caption{Theoretical results for resistivity as a function of temperature for
parameters of sample B showing a zoom in of the low-temperature region
in (b) of the results in panel (a). (c) Red circles and black squares
correspond to numerical calculation of the minimum and maximum in
the temperature dependence of resistivity for sample B, respectively. }
\label{fig:SampleBTheory}

\end{figure}

In Fig.~\ref{fig:SampleBTheory}, we present the theoretical calculation
for the parameters corresponding to the experimental sample B, which
qualitatively simulates the experimental results for sample B shown
in Figs.~\ref{rhoT}(b),~\ref{rhoT}(d),~and~\ref{MinMax}(b). Comparison of Figs.~\ref{rhoT}~and~\ref{MinMax}
with Figs.~\ref{rhoTtheory}-\ref{fig:SampleBTheory} establishes
that the crossover picture of 2D MIT is valid qualitatively (and probably
even quantitatively), i.e., both the screening-induced metallic temperature
dependence and the quantum weak localization temperature dependence
are present in the resistivity.

\section{Discussion and conclusion}

In this paper, we present experimental and theoretical results for
the density-dependent low-temperature transport properties of high-mobility
2D Si MOSFETs manifesting the 2D MIT phenomena with the critical goal
of searching for the possible presence of the quantum weak localization
(i.e., $\ln T$) corrections to the resistivity on the apparent metallic
side of the so-called 2D metal-insulator transition. Our detailed
analyses of the experimental transport data indicate the existence
of a resistivity minimum at a density-dependent characteristic low
temperature in the effective metallic regime. The existence of this
characteristic temperature, with resistivity $\rho(T)$ increasing
with both increasing and decreasing temperature away from this minimum,
points to the presence of two competing transport mechanisms in the
system with one being ``metallic'' (i.e., $d\rho/dT>0$) and the
other ``insulating'' (i.e., $d\rho/dT<0$) which balance each other
at this low-$T$ resistivity minimum with the localization effect
dominating at still lower temperatures. We identify the two competing
mechanisms to be the temperature-induced reduction of screening (leading
to the metallic $d\rho/dT>0$ behavior) and the quantum interference
induced weak localization contribution (leading to the $\ln T$ weak
localization correction with $d\rho/dT<0$) which dominate, respectively,
the higher and the lower temperature sides of the resistivity minima.
We find that the minimum occurs mostly at temperatures which are experimentally
inaccessible in high-mobility samples (due to perhaps the well-known
electron heating problem in 2DEG semiconductors), thus providing a
possible explanation for why most experimental measurements in high-mobility
samples do not manifestly show the $\ln T$ insulating behavior at
low temperatures in the metallic regime. We believe that the $\ln T$
weak localization behavior would routinely show up in the metallic
transport properties of high-mobility samples if much lower experimental
electron temperatures can be achieved in future measurements. In fact,
our work indicates that the most straightforward experimental technique
to search for signatures of localization in the metallic regime of
high-mobility 2D semiconductor systems is to look for extrema in the
resistivity $\rho(T)$ at a fixed density $n$ by numerically obtaining
the solutions of $d\rho/dT=0$ in the experimental $\rho(T,n)$ data
at fixed density. The low-temperature minima in the resistivity would
correspond to the temperature at each density below which localization
correction dominates the semi-classical metallic temperature dependence.
Depending on the carrier density, this minimum could lie at inaccessibly
low temperature, but there still should be some signatures for the
minima in the data. As observation of this low-temperature density-dependent
 minima (Fig.~\ref{MinMax} in our samples) in the transport
data indicates that the 2D MIT is a crossover and not a true quantum
phase transition, and the absence of a manifest $\ln T$ weak localization
effect in the resistivity is simply a feature of the insulating localization
correction being overwhelmed by a strong metallic temperature dependence
in the semi-classical Drude-Boltzmann resistivity as our theoretical
results in Figs.~\ref{rhoTtheory}-\ref{fig:SampleBTheory} clearly
demonstrate. The excellent qualitative agreement between our theory
and our experiment is a strong evidence in favor of 2D MIT being a
crossover phenomenon. The possibility of 2D MIT being a Fermi-liquid
crossover phenomenon driven by disorder in high-mobility low-density
MOSFETs with weak localization effects masked by finite-temperature
Drude-Boltzmann effects was also pointed out in the early experimental
works of Pudalov~\cite{Altshuler_Martin_Maslov_Pudalov_Prinz_Brunthaler_Bauer_2000,Altshuler_Maslov_Pudalov_2000,Prus_2001}
and of Pepper~\cite{Lewalle_2002,Lewalle_2004}.

One feature, prominent both in our experimental data Figs.~\ref{rhoT}~and~\ref{MinMax}
and in our theory (Figs.~\ref{rhoTtheory}-\ref{fig:SampleBTheory})
needs to be specifically mentioned in addition to the resistivity
minima discussed before. It is the existence of the high-temperature
resistivity maxima in the data (black squares in the figures) with
$d\rho/dT<0$ above this temperature (until phonons become important
at still higher temperatures). This quantum-classical high-temperature
crossover behavior is ubiquitous in all high-mobility 2D systems in
the metallic phase, where after the sharp initial rise of $\rho(T)$
with increasing $T$, $\rho(T)$ goes through a maximum at a density-dependent
``high'' temperature slowly decreasing beyond this characteristic
temperature until phonon scattering takes over at still higher temperatures.
We note that the characteristic temperature for this resistivity maxima
(black squares) rapidly decreases with decreasing density, whereas
the characteristic temperature for the resistivity minima (red circles)
increases with decreasing density. Once these two lines come close
together ($n\sim1.2\times10^{11}\mathrm{cm}^{-2}$ for sample A and
$n\sim1.8\times10^{11}\mathrm{cm}^{-2}$ for sample B, see Fig.~\ref{MinMax}),
the system simply behaves as an insulating system at all lower densities
since $d\rho/dT<0$ for all density and temperature below this intersection
regime of the black squares and red circles. This finite-temperature
behavior is also apparent in our theoretical curves [see Fig.~\ref{fig:MinMaxTheory}
for sample A and Fig.~\ref{fig:SampleBTheory}(c) for sample B]. 

Before concluding, we point out that, within our model of Boltzmann
resistivity due to screened charged impurity scattering and weak localization
due to quantum interference [i.e., Eq.~(\ref{eq:Resistivity}) in
Sec.~\ref{sec:Temperature-dependence-of}], we can actually derive
a leading order analytical formula for the characteristic temperature
(i.e., the red circle plots in the figures) for the resistivity minima
below which weak localization effect should dominate the metallic
transport properties. Using the leading order (linear) analytical
low-temperature expansion in temperature for $\rho_{B}(T)$ in Eq.~(\ref{eq:Resistivity})
we get for the characteristic temperature $T_{m}$ where $d\rho/dT=0$
to be,

\begin{equation}
T_{m}\propto T_{F}/\sigma_{B}(T=0),
\end{equation}
where $T_{F}\propto n$ is the Fermi temperature and $\sigma_{B}(T=0)=\rho^{-1}(T=0)$
is the zero-temperature Boltzmann conductivity due to charged impurity
scattering. It is well-known~\cite{DasSarma_2013b} that $\sigma_{B}(T=0)$
obeys an approximate scaling law with the carrier density going as,

\begin{equation}
\sigma_{B}\sim n^{\alpha+1},
\end{equation}
where $\alpha(n)\approx0.3$ in Si MOSFETs in the low-density metallic
regime. This leads to a rather weak density dependence of the characteristic
temperature $T_{m}$ going as,

\begin{equation}
T_{m}\sim n^{-0.3},
\end{equation}
which is approximately consistent with the experimental and theoretical
numerical results for the red circle lines in Figs.~\ref{MinMax},~\ref{fig:MinMaxTheory},~and~\ref{fig:SampleBTheory}(c).
The important point to note is that the weak localization correction
becomes important at lower density since the weak localization effect
becomes progressively stronger with the decreasing Drude conductivity
with decreasing carrier density. On the other hand, if $\sigma_{B}(T)$
is temperature independent as it is for low-mobility samples, the
weak localization $\ln T$ correction would be visible at all carrier
densities.

We conclude by emphasizing that our theory has many approximations
which need to be improved in future work. We neglect effects of interaction
in the theory beyond the finite-temperature screening effect by the
electrons themselves which we include within RPA (i.e., infinite sum
of bubble diagrams). At the low carrier densities of interest in the
2D MIT problem, interaction effects are likely to be important, but
we neglect them in the spirit of obtaining the leading-order result
within the minimal model. In addition, we believe that the interaction
effects may be substantially suppressed by finite temperature since
$T/T_{F}$ is not particularly small at the experimental densities
and temperatures. We also assume rather unrealistically that the weak
localization quantum correction may simply be added to the Drude-Boltzmann
conductivity as a $\ln T$ correction, which is obviously a simplification
done in the spirit of developing the minimal physical model for including
both metallic and insulating temperature dependence within a single
unifying scheme. In particular, neither the Boltzmann theory nor the
simple weak localization correction is strictly applicable in the
strongly disordered situation close to the metal-insulator transition
where $k_{F}\ell\sim1$, but we have assumed in this work that
such a minimal theory (i.e., Boltzmann conductivity along with the
$\ln T$ weak localization quantum correction) can be continuously
extended from the high-density ($k_{F}\ell\gg1$) regime to the low-density
regime ($k_{F}\ell\gtrsim1$) as long as the system is still nominally
in the metallic phase. Our minimal theory obviously becomes progressively
quantitatively worse as the carrier density decreases, but we think
that it remains qualitatively valid all the way down to $k_{F}\ell\gtrsim1$
in the metallic phase. We have neglected all phonon scattering effects
in the theory which probably become important for $T\gtrsim10\mathrm{K}$
outside the regime of our interest. It is straightforward to include
phonon scattering in the theory and it adds a resistivity increasing
linearly with $T$ for $T>T_{\mathrm{BG}}$ ($\sim10\mathrm{K}$ in Si MOSFET)
where $T_{\mathrm{BG}}$ is the Bloch-Gruneisen temperature. Since our interest
in the current paper is the low-temperature 2D MIT physics, phonon
scattering effects are irrelevant for our problem. A 2DEG in the presence of disorder 
 and electron-electron interaction is expected to manifest a diffusive (Altshuler-Aronov) interaction correction to conductivity~\cite{AAbook,AkkermansBook,Zala_2001,Zala_2001R,Gornyi_2004,Burdis_1988,Klimov_2008} which gives another(i.e. in addition to the weak localization correction) $\ln T$ contribution in the temperature dependence of conductivity. However, the prefactor in front of $\ln g T$ given by a combination of singlet and 
triplet components is not known at low electron densities of interest in the current work since interaction effects are non-perturbatively strong at low carrier densities. This Altshuler-Aronov effect does not change the functional form of the temperature dependence of resistivity, and therefore our theory as described by Eqs.~(\ref{eq:Resistivity})~and~(\ref{eq:WLT0}) not including the Altshuler-Aronov part of the electron-electron interaction correction (note that the ballistic part of the electron-electron interaction effect is included in our Boltzmann theory of Sec.~\ref{sec:Boltzmann}) nevertheless qualitatively describes the data. It is, however, important to point out that naively adding an Altshuler-Aronov lnT term in our theory will be an incorrect double-counting of many-body effects since the screening effect we included non-perturbatively in the theory already contains the Hartree part of the interaction effect in the ballistic regime (which crosses over to the lnT effect in the diffusive regime~\cite{Zala_2001,Zala_2001R},  and thus there is no need to add a separate lnT term arising from the Altshuler-Aronov effect also.

Finally, we mention that we have not discussed at all the nature of
the actual crossover to the strongly localized exponential temperature
dependence (in the resistivity) at very low carrier density as our
focus in this work has entirely been on the signature of weak
localization in the putative metallic regime. The issue of the strong
localization crossover has recently been discussed in great detail
by the two of us~\cite{DasSarma_2014}. One key issue that remains
open in this context is the role of impurity-induced density inhomogeneity
or puddle formation in the 2DEG as the system crosses over to the
strongly insulating phase and screening fails completely. Such inhomogeneous
puddles could lead to percolation physics competing with the physics
of Anderson localization. In fact, sometimes the crossover to the
strong localization behavior may itself be considered a percolation
transition as was done in Ref.~\onlinecite{Tracy_2009}. The interplay
of puddle physics and localization physics in the strongly interacting
2D system is an interesting open question in the 2D MIT problem. For
our specific considerations, the puddle size could act as a cut off
for the dephasing length explaining why the low-temperature dephasing
length appears to be short compared with the standard Fermi liquid
theory.

This work is supported by NSA-LPS-CMTC.

\end{document}